\documentclass[aps,prl,epsfigure,twocolumn,showpacs,superscriptaddress]{revtex4}

\usepackage{graphics}
\usepackage{epsfig}

\begin{document}
\title{On the quantum criticality in the ground and the
thermal states of XX model
}

\author{W. Son}
\affiliation{Centre for Quantum Technologies, National
University of Singapore, 3 Science Drive 2, Singapore 117543 }

\author{V. Vedral}
\affiliation{Centre for Quantum Technologies, National
University of Singapore, 3 Science Drive 2, Singapore 117543}
\affiliation{\text{The School of Physics and Astronomy,
University of Leeds, Leeds, LS2 9JT, United Kingdom}}
\affiliation{Department of Physics, National University of Singapore,
2 Science Drive 3, Singapore 117542}
\date{\today}

\begin{abstract}
We compare the critical behavior of the ground state and
the thermal state of the XX model. We analyze the full energy
spectrum and the eigenstates to reconstruct the ground state
and the thermally excited state. With the solutions, we discuss
about several physical properties of the states, which are
related to quantum phase transition, in various limits, at zero
temperature as well as at a thermal equilibrium.
\end{abstract}

\maketitle

Conventional definition of quantum phase transition (QPT) is given
by energy level crossing of ground energy state with varying
external parameter \cite{Sachdev99}. The definition is based on
the view that the ground energy state is taken as a {\it quantum
phase} itself and, thus, QPT is a transition occurring at the point where a zero temperature state is transited from one state to the other.
Because of the singular structure of the energy transition, the
point of QPT is known to be detected by non-analyticity of free
energy.

Within the definition, a difficulty may arise if one tries to
consider a finite temperature QPT. It is because QPT is only
defined at $T=0$. Typical treatment of finite temperature phase
transition so far is provided by the non-analyticity of
thermodynamic properties or the characteristic wave length of
the system \cite{Chakravarty89,Sachdev97}. However, it is not
straightforward how the singularity in the thermal state is
consistently combined with the notion of energy level crossing
at zero temperature.

In this article, we study the full solutions of XX spin chain
model, not only the energy eignevalues but also the eignestates in the original spin basis. With the solutions, we
demonstrate how the sequential energy level crossings at zero
temperature is led to a continuous QPT in infinite spin limit,
$N\rightarrow \infty$. In the XX spin model, the number of
ground energy level-crossings grow with the size of the chain.
When $N\rightarrow \infty$, the crossing points become dense
within a sharply defined critical region, giving rise to a
continuous QPT. This is a typical example when a ground state is
no longer in a {\it pure state} but in a {\it mixed state}. At the same time, each crossing point of the ground energy state becomes fully
analytic.

The situation is comparable to a state in finite temperature
which is a mixture of all the state spectrums with the
Boltzmann distribution. We investigate the structure of
state in a thermally excited XX spin chain and the
purity of the state to compare it with the ground state of the
infinite chain. In the case, it can be seen that the temperature
is acting as a controlling parameter to determine the degree of
state mixture in the system. Thermodynamic properties of the
thermal state are remained analytic in the region. As temperature is increased, the state moves into a thermal mixture which destroys the quantum correlations,
entanglement, in the system. Extensive discussions on the
relation between QPT and entanglement in the XX model can also
be found at \cite{Son08,Amico07,Hide07}.

\section{XX Model and full solutions}

\begin{figure}[t]
\begin{center}
{\bf (a)}\hskip3.5cm{\bf (b)}
\centerline{
\includegraphics[width=1.5in]{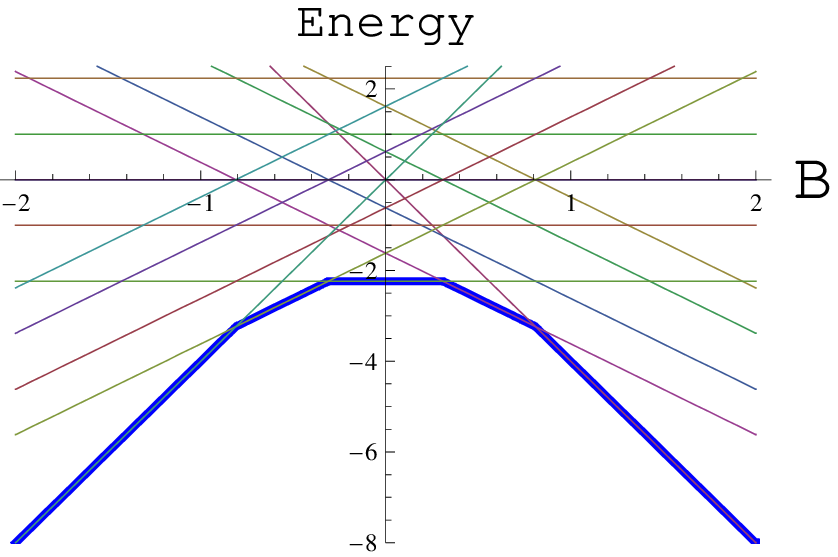}
\includegraphics[width=1.5in]{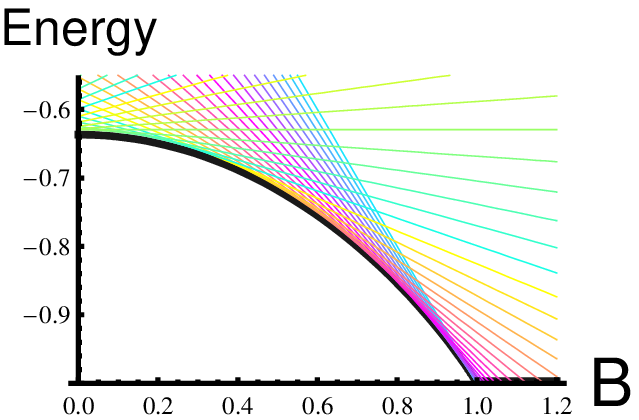}}
\end{center}
\caption{(a) The spectrum of $2^4$ energy levels
for the $N=4$ XX spin chains. The energies vary
as a function of magnetic field $B$. Blue lines
are the energy change of the ground state.
Energy level crossings occur at the critical
points in the ground state as well as in excited states.
Ground state entanglement is jumping as the state
is switched from an entangled state to another state.
(b) The ground state energy per spin against the
magnetic field $B$ in the thermodynamic limit.
The energy is plotted together with large spin
chain case, $N=50$. The continuous energy level
crossings occur in the limit at the region
$0<B<1$.
}
\label{fig:Infinit}
\end{figure}

We start our discussion by considering $N$ spin $1/2$ particles
in a line, coupled by nearest neighbor XX interaction, with
Hamiltonian
\begin{equation}
H=-\Big[\sum_{i=1}^{N} \frac{J}{2}(\sigma_{i}^{x}\sigma_{i+1}^{x}
+\sigma_{i}^{y}\sigma_{i+1}^{y}) +B\sigma_{i}^{z}\Big]
\end{equation}
where we have assumed that the exchange interaction $J$ fixes
the energy unit. The model with periodic boundary condition has
been firstly solved by Katsura \cite{Katsura62} and recently
discussed at finite size in Ref. \cite{dep}. Here, we assume
open boundaries, with $\sigma_{N+1}=0$. After Jordan-Wigner (JW)
and Fourier transformations,
\begin{equation}
\label{eq:transform}
d_k =\sqrt{\frac{2}{N+1}} \sum_{l=1}^{N}
\sin\left(\frac{\pi k l}{N+1}\right) \, \bigotimes_{m=1}^{l-1}
\sigma_m^z \sigma_l^-,
\end{equation}
the Hamiltonian takes a diagonal form, $H = \sum_{k=1}^{N}
\Lambda_k d_k^{\dagger}d_k + N B$ where $\Lambda_k=-2B +
2\cos\left[(\pi k)/(N+1)\right]$. The $2^N$ eigenenergies are
$\epsilon_l \equiv\sum_{k=1}^{N} \Lambda_k \langle \psi_l
|d_k^{\dagger}d_k|\psi_l\rangle + N B$ and the corresponding
eigenstates are found
\begin{equation}
|\psi_l\rangle = \prod_{k=1}^N (d^{\dagger}_k)^{\alpha_k^{(l)}}
\, |\Omega\rangle,
\end{equation}
where the binary integer number $\alpha_k^{(l)}\in \{0,1\}$
identifies eigenenergy as $\epsilon_l \equiv\sum_{k=1}^{N}
\Lambda_k\alpha_k^{(l)} + N B$. The state $|\Omega\rangle$ is
the vacuum: $d_k|\Omega\rangle=0 \, \forall k$. It is also
useful to know that $\langle\psi_i|\psi_j\rangle=\delta_{i,j}$.
The energy diagram of $\epsilon_l$ for $N=4$ as a function of
$B$ is illustrated in Fig.\ref{fig:Infinit}(a).

%To view the state in a simpler way, it is convenient to define
%the binary numbers $\alpha_k^{(l)}$ in terms of a $N$
%dimensional orthogonal binary vectors
%\begin{equation}
%\vec{\alpha}_{l}= \sum_{k=1}^{N}\alpha_k^{(l)} \hat{v}_k
%\end{equation}
%where $\hat{v}_k$ are orthonormal unit vectors satisfying
%$\hat{v}_k\cdot\hat{v}_{k'}=\delta_{k,k'}$. There exist $2^N$
%different vectors $\vec{\alpha}_{l}$

\section{Ground state of XX spin chain}
In this section, we discuss about the ground state of the XX
model. The ground energy state is the lowest energy state which
is a function of the transverse magnetic field $B$. Letting
$B_k=\cos[k \pi/(N+1)]$ and defining the $k$-th region, $B_{k+1}
< B < B_k$, we can find the ground state $|\psi_g^k\rangle= d_k
d_{k-1} \cdot\cdot\cdot d_1|\psi_g^0\rangle=\prod_{l=k+1}^{N}
d_l^{\dagger}|\Omega\rangle$ and its energy  as $ \epsilon_g^k =
-(N-2k) B -2 \sum_{l=1}^{k} \cos\left(\frac{\pi l}{N+1}\right)
$ (refer \cite{Son08} for detail). The index $k$ is an integer in
$1\leq k<N$ and $B_k$ represents the points where energy level
crossings occur. When $B<B_N$ or $B>B_1$, no other crossing
exists and the ground state is simply given by the fermion
vacuum state. The ground state energy for $N=4$ is plotted in
Fig. \ref{fig:Infinit} (a) by blue lines.

The ground state in the spin bases after $k$ level crossing can
be obtained recursively
$|\varphi_g^k\rangle=\prod_{k'=1}^{k}\left[\sum_{l=1}^{N}
S_l^{k'}~ \left(
\prod_{m=1}^{l-1}\sigma^z_m\sigma_l^{-}\right)\right]
|\uparrow\rangle^{\otimes N}$ where
$S_l^k\equiv\sqrt{2/(N+1)}\sin\left[(\pi k l)/(N+1)\right]$ and
the explicit form of the state is given as
\begin{equation}
|\varphi_g^k\rangle=\left(\frac{2}{N+1}\right)^{\frac{k}{2}}
\sum_{l_1 < l_2 <\cdot\cdot\cdot <l_k} C_{l_1 l_2
\cdot\cdot\cdot
l_k} |l_1, l_2, \cdots, l_k\rangle
\end{equation}
where $|l_1, l_2, \cdots, l_k\rangle$ is the state whose
$l_1$-th, $l_2$-th,$\cdots$, $l_k$-th indicate the position of
flipped spin. The amplitudes in the state %(\ref{eq:state})
are $C_{l_1 l_2 \cdot\cdot\cdot l_k} =\sum_P (-1)^P
\sin\left(\frac{\pi P(1) l_1}{N+1}\right)\sin\left(\frac{\pi
P(2) l_2}{N+1}\right)\cdots\sin\left(\frac{\pi P(k)
l_k}{N+1}\right)$. Here, the sum extends over all the
permutation of the numbers from 1 to k, (denoted by $P(k)$). We
point out that, at each crossing point (caused by the variation
of $B$), the ground state jumps discontinuously from one
symmetric subspace to the another, orthogonal to the previous
one. To see the state explicitly, we illustrate the ground
states for $N=4$ in spin bases which are
\begin{widetext}
\begin{eqnarray}
|\varphi_g^0\rangle&=&|\uparrow,\uparrow,\uparrow,
\uparrow\rangle,
~~~~
|\varphi_g^1\rangle=a_1^-
 |\downarrow,\uparrow,\uparrow,\uparrow\rangle
+a_1^+|\uparrow,\downarrow,\uparrow,\uparrow\rangle
+a_1^+|\uparrow,\uparrow,\downarrow,\uparrow\rangle
+a_1^-|\uparrow,\uparrow,\uparrow,\downarrow\rangle\nonumber\\
|\varphi_g^2\rangle&=&
a_2\left( |\downarrow,\downarrow,\uparrow,\uparrow\rangle
+2|\uparrow,\downarrow,\downarrow,\uparrow\rangle
+2|\downarrow,\uparrow,\uparrow,\downarrow\rangle
+|\uparrow,\uparrow,\downarrow,\downarrow\rangle
+\sqrt{5}|\downarrow,\uparrow,\downarrow,\uparrow\rangle
+\sqrt{5}|\uparrow,\downarrow,\uparrow,\downarrow\rangle\right)
\\
|\varphi_g^3\rangle&=&
 a_3^-|\uparrow,\downarrow,\downarrow,\downarrow\rangle
+a_3^+|\downarrow,\uparrow,\downarrow,\downarrow\rangle
+a_3^+|\downarrow,\downarrow,\uparrow,\downarrow\rangle
+a_3^-|\downarrow,\downarrow,\downarrow,\uparrow\rangle,~~~~
|\varphi_g^4\rangle=
|\downarrow,\downarrow,\downarrow,\downarrow\rangle\nonumber
\end{eqnarray}
\end{widetext}
where
$a_1^{\pm}=-a_3^{\pm}=\frac{1}{2}\sqrt{1\pm\frac{1}{\sqrt{5}}}$
and $a_2=-\frac{1}{2\sqrt{5}}$. From the state, it is clear that
the superscript of $\varphi$ indicates the number of down spins.
The ground state at the $k$-th region is composed of
$_NC_k=N!/[k!(N-k)!]$ orthogonal vectors in spin bases which are the
possible choices of k spins, out of total N spins. The ground states
are in an orthogonal subspace, $\langle
\varphi_g^k|\varphi_g^{k'}\rangle=\delta_{k,k'}$.

It is found that the ground energy spectrum and the ground
state become continuous at the thermodynamic limit,
$N\rightarrow\infty$. In the limit, the sum in $\epsilon_g^k$
turns into a definite integral and thus the ground state energy
reads
\begin{equation}
\displaystyle
\lim_{N\rightarrow\infty}\frac{\epsilon_g(B)}{N}
=\frac{2}{\pi} \left[ B\left(\arccos B-\frac{\pi}{2}\right)
-\sqrt{1-B^2} \right].
\end{equation}
Within the critical region the ground state energy is analytic
everywhere except for $B=\pm 1$. From the finite size analysis
we observe that such a critical line becomes a dense set of
level crossing (see Fig. \ref{fig:Infinit} (b)) and therefore is
possible to be considered as a line of continuous QPT. At the
crossing points $B_k$, the state exists as a mixed state with
equal mixture of the states in a neighboring $k$-th region,
$|\varphi_g^k\rangle$ and $|\varphi_g^{k+1}\rangle$, {\it i.e.}
\begin{equation}
\label{eq:thermostate}
\rho(B_k)=\frac{1}{2}\left(|\varphi_g^k\rangle\langle
\varphi_g^k|+|\varphi_g^{k+1}\rangle\langle
\varphi_g^{k+1}|\right).
\end{equation}
Therefore, one can find that the state in the thermodynamics
limit is also varying from a mixed state into the other mixed
state continuously as a function of $B$. That is because the
critical points $B_k$ is now continuous function,
$B_{\omega}=\lim_{N\rightarrow\infty}\cos (k\pi/(N+1))=\cos
(\pi\omega)$ where $0<\omega<1$. Moreover, one would find that
the mixedness of the state in the limit is remained as a
constant for any $B$ in $-1<B<1$ as $\mbox{Tr}\rho(B)^2=1/2$.
The mixedness (or purity) $1/2$ of the ground state in the limit
can also be proved from the limiting case of a thermal state
which will be discussed in the following section.

\section{XX Spin chain in thermal equilibrium}

In this section, we investigate the state in a thermal
equilibrium and discuss about the properties related with
quantum phase transition. At a finite temperature, the state
exists in a mixture of all the energy state distributed by
Boltzmann statistics. The state in a diagonal basis
$|\psi_l\rangle = \Pi_{k=1}^N (d^{\dagger}_k)^{\alpha_k^{(l)}}
\, |\Omega\rangle$, $\alpha_k^{(l)}\in \{0,1\}$ is given as
\begin{equation}
\label{eq:thermal}
\rho_T=\sum_{l=1}^{2^N} p_l|\psi_l\rangle\langle\psi_l|
\end{equation}
where $p_l=\exp(-\beta \epsilon_l)/Z$ is the Boltzmann
distribution of the thermal state with the partition function
$Z=e^{-\beta
NB}\prod_{k=1}^{N}\left(1+e^{-\beta\Lambda_k}\right)$. In terms
of the spin basis, the thermal state further can be decomposed
into the states which are in a symmetric subspace as,
\begin{equation}
\rho_T=\sum_{m=0}^{N} \sum_{r=1}^{_NC_m}
q_{r}^{m}|\varphi_{r}^{m}\rangle\langle\varphi_{r}^{m}|.
\end{equation}
The index $m$ in the superscript indicates a symmetric subspace
where vector components of the state contain the $m$ number of
spins flipped and the subscript index $r$ identifies the
${_NC_m}=N!/[m!(N-m)!]$ orthogonal vectors within the subspace.
Thus, the state satisfies the orthonormal condition,
$\langle\varphi_r^m|\varphi_{r'}^{m'}\rangle
=\delta_{r',r}\delta_{m',m}$.
The state $|\varphi_{r}^{m}\rangle$ can be obtained from the
$l$-th state $|\psi_l\rangle$ using the transformation in
(\ref{eq:transform}),
%\begin{equation}
$|\varphi_r^m\rangle=\prod_{k=1}^{N}\left[\sum_{j=1}^{N}
S_j^{k}~ \left(
\prod_{i=1}^{j-1}\sigma^z_i\sigma_j^{-}\right)\right]^{\alpha_k^{(l)}}
|\uparrow\rangle^{\otimes N}
$%\end{equation}
where $S_j^k\equiv\sqrt{2/(N+1)}\sin\left[(\pi j k
)/(N+1)\right]$ and $\alpha_k^{(l)}\in \{0,1\}$. The values of
the label $l$ for the binary vectors
$\vec{\alpha}_l=\sum_k\alpha_k^{(l)}\hat{v}_k$ are determined by
$r$ and $m$ through an index match $l=r+\sum_{s=0}^{m-1}{_NC_s}$
under the constraints $\sum_{k=1}^{N}\alpha_{k}^{(r)}= m$. (It
starts from $l=1$ when $(r,m)=(1,0)$.) Similarly, the
probabilities of the Boltzmann distribution are given as
$p_l\equiv q_{r}^{m}=\exp(-\beta \epsilon_r^m)/Z$. The lowest energy state for a fixed $m$
become a ground state at a given region of $B$, {\it i.e}
$|\varphi_1^m\rangle\equiv |\varphi_g^m\rangle$. Furthermore,
the corresponding energy eigenvalues by the subspace indexes $m$
and $r$ can be found as
\begin{equation}
\epsilon_l\equiv\epsilon_r^m=-(N-2m)B-2\sum_{k=1}^N \cos[\pi
k/(N+1)]\alpha_k^{(r)}.
\end{equation}
This shows that the slop of the energy against $B$ is invariant
for the states in a given symmetric subspace, with a fixed $m$,
and, thus, it also gives a proof that {\it the energy level
crossings of excited states occur only between the states in a
different symmetric subspace}. The level crossings of the
excited states for $N=4$ spins also can be seen in Fig.
\ref{fig:Infinit} (a).

\begin{figure}[h]
\begin{center}
{\bf (a)}\hskip3.5cm{\bf (b)}
\centerline{
\includegraphics[width=1.5in]{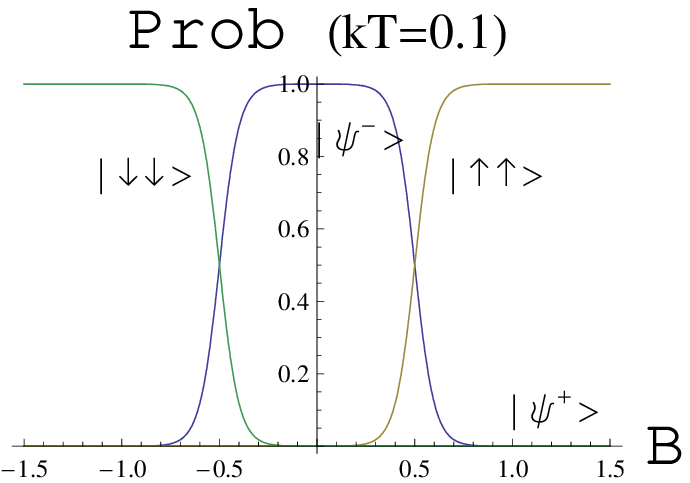}
\includegraphics[width=1.5in]{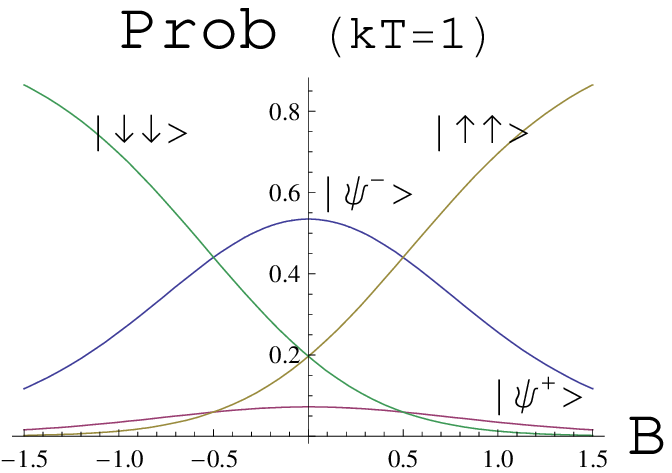}
}
\end{center}
\caption{Distributions of thermal state population for the
$N=2$ XX spin chain when (a) $kT=0.1$ and (b) $kT=1$.
The states $|\psi^-\rangle$ and $|\psi^+\rangle$
are the singlet and the triplet. Critical behavior
disappears as the system goes into
a thermal equilibrium.}
\label{fig:Temp}
\end{figure}

\begin{figure}[h]
\begin{center}
{\bf (a)}\hskip3.5cm{\bf (b)}
\centerline{
\includegraphics[width=1.5in]{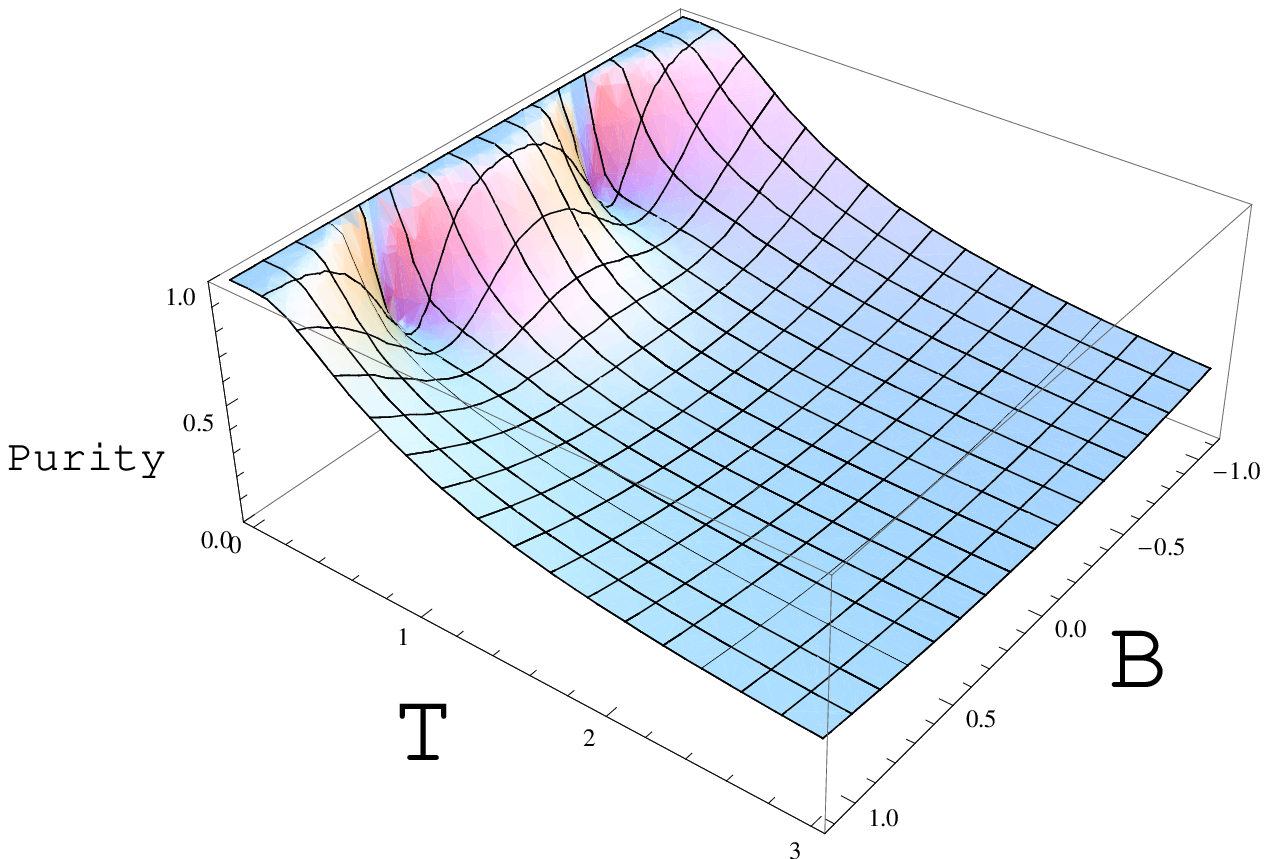}
\includegraphics[width=1.5in]{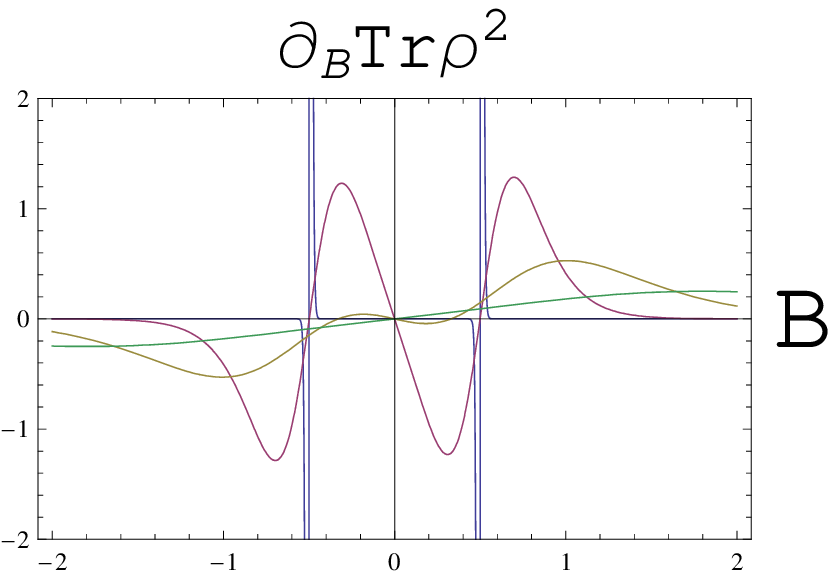}
}
\end{center}
\caption{(a) Purity of two qubit system
as a function of magnetic field $B$ and temperature $T$.
(b) Derivatives of the purity as a function of $B$ at
different temperatures, $T=0.01$ (blue), $T=0.3$ (pink),
$T=0.8$ (yellow) and $T=2$ (green). At a low temperature
region, the purity of the system shows peaks near the
critical points $B =\pm 1/2$.
This is the point where the derivative of purity w.r.t. B
becomes singular when $T=0$. At the high temperature region,
the purity becomes monotonic convex function of B and the derivative
of the purity approaches to a linear function.}
\label{fig:purity2}
\end{figure}

As a special case, we inspect the density matrix of the
thermally excited state when $N=2$ which is in a form,
\begin{widetext}
\begin{equation}
\rho_T^{N=2}=p_1 |\uparrow,\uparrow\rangle\langle
\uparrow,\uparrow|
+p_2|\psi^-\rangle\langle\psi^-|
+p_3|\psi^+\rangle\langle\psi^+|
+ p_4 |\downarrow,\downarrow\rangle\langle
\downarrow,\downarrow|
\end{equation}
\end{widetext}
where $|\psi^{\pm}\rangle=\frac{1}{\sqrt{2}}
\left(|\uparrow,\downarrow\rangle
\pm|\downarrow,\uparrow\rangle\right)$. The state is expended in
the three different symmetric subspaces, where there is no spin
flip (m=0), one spin flip (m=1) and two spin flips (m=2). Each
subspace contains $_2C_0$, $_2C_1$ and $_2C_2$ orthogonal
elements of states, $\{|\uparrow,\uparrow\rangle\}$,
$\{|\psi^-\rangle,|\psi^+\rangle\}$ and
$\{|\downarrow,\downarrow\rangle\}$. We plot the probability
distribution of the states $p_l$ in Fig.\ref{fig:Temp}. One
would find that the distribution and the partition function
become analytic for any $B$ and $T$ as $T$ is away from zero. If
we investigate entanglement in the system, one would identify
that the state become separable when $4p_1 p_4 > (p_2-p_3)^2$,
(from the negativity \cite{Peres}, the negative eigenvalue of
partially transposed state), leading to a critical condition for
temperature $kT>1.13459$. Interestingly, the separability
condition for the two qubit thermal state is independent of the
strength of magnetic field $B$ {\it except when $T=0$}. Thus,
the singular behavior of entanglement with respect to $B$ is
washed out as the temperature increased. This also has been
depicted by the purity of the two qubit system in Fig.
\ref{fig:purity2}.

\begin{figure}[th]
\begin{center}
\centerline{
\includegraphics[width=2.3in]{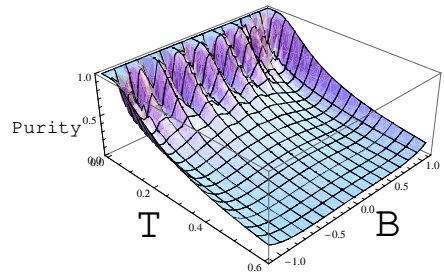}
}
\end{center}
\caption{Plot of the purity as a function of $B$ and $T$
when $N=10$. The state is in a pure state
when $\mbox{Tr}\rho^2=1$ and is in a completely mixed
state when $\mbox{Tr}\rho^2=1/2^{10}$. The
purity of state is singular at the critical points of
$B$ where the state exists as a mixture of
two orthogonal symmetric states. It becomes smaller
as $T$ is increased.}
\label{fig:purity}
\end{figure}

With the density matrix of the thermal state (\ref{eq:thermal}),
we investigate a physical property which may be directly related
to the critical behavior of the system. The mixedness of the
thermal state for any $N$ is
\begin{equation}
\mbox{Tr}\rho_T^2=\sum_l p_l^2=
\prod_{k=1}^{N}\Big[1-(1+\cosh\beta\Lambda_k)^{-1}\Big].
\end{equation}
This characterizes how far the state is from a pure state. The
purity is plotted in Fig.\ref{fig:purity2} for a small number of
qubits ($N=2$) and for a large number of qubits ($N=10$) in
Fig.\ref{fig:purity}. At the $T\rightarrow\infty$ limit
($\beta\rightarrow 0$), the purity becomes $1/2^N$. Thus, in the
high temperature region and the thermodynamic limit,
$N\rightarrow \infty$, the purity approaches to zero where the
state is in a complete mixture of infinitely many orthogonal
states. At the opposite extreme, $T\rightarrow 0$, the purity
becomes either $1$ or $1/2$ depending on the $B$ value. It is $1/2$
when $B$ is $\cos[k\pi/(N+1)]$ and 1 otherwise. When
$N\rightarrow \infty$, the purity becomes $1/2$, {\it i.e.}
$\lim_{N\rightarrow \infty}\lim_{T\rightarrow
0}\mbox{Tr}\rho_T^2=1/2$ in $-1<B<1$ because there exits at
least a single value of $B$ in the interval to make
$\Lambda_k=0$ $\exists k$. In the $|B|>1$ region, the purity is
1. This proves that the ground state of the infinite XX chain in
the region $|B|<1$ is the equal mixture of the neighboring
states as (\ref{eq:thermostate}). Furthermore, it is
straightforward to see that entanglement disappears in the
region $|B|>1$ and at high temperature. The separability of
infinite XX chain with entanglement witness has been fully treated in
\cite{Hide07}.

In our analysis of the XX spin chain, it has been shown that the
fundamental difference between the thermal state and the ground
state of the infinite chain is only the degree of state
mixedness. In both of the cases, the changes of states are
impossible to be detected by singularity due to the continuity
of state changes. In that sense, it can be pointed out that the
energy level crossing is a particular mode of the ground state change
when the state is in {\it a pure state}. In general, the state change can be detected by the change of {\it entanglement}, up to local unitary, which is destroyed by mixedness of the states.  In fact, entanglement identifies the true quantum phase (or state)
of a system since it is the property which can
exist only in quantum systems whether they are pure
or mixed \cite{Werner89}. Our approaches are general
enough to be used for the discussions about the QPT and related properties of any other many particle systems.

%\begin{itemize}
%\item What is the entanglement of the excited state?
%(The ground %energy
%    state is in a
%    symmetric subspace. Are the excited states in the symmetric
%    subspace too??)
%\item Are there any critical temperature where
%entanglement %disappears?
%    Can the critical
%    temperature be obtained by analytically?
%\item Can we make an analysis of disappearance of
%entanglement %in terms
%    of mixedness?
%    (Relation between the criticality and mixedness or
%purity...)
%\item Can I calculate some properties which have never been
%calculated
%    exactly before from the
%    properties of symmetry?
%\end{itemize}
%%%
%%\subsection{Block Entropy}

{\it Acknowledgement}- This work is supported by the National
Research Foundation \& Ministry of Education, Singapore.

\end{document}